\documentclass[aps,prl, twocolumn,
               10pt,
               longbibliography,
               floatfix,
               nofootinbib,
               superscriptaddress]{revtex4-2}
\usepackage{cancel}
\usepackage{physics}
\usepackage[english]{babel}
\usepackage[utf8]{inputenc}
\usepackage{amssymb}
\usepackage{amsmath}
\usepackage{multirow}
\usepackage{float}
\usepackage[pdftex]{graphicx}
\usepackage{xspace}
\usepackage{dsfont}
\usepackage{soul}
\usepackage{bm}
\usepackage[normalem]{ulem}
\usepackage{hyperref}

\hypersetup{
    colorlinks,%
    citecolor=blue,%
    filecolor=blue,%
    linkcolor=blue,%
    urlcolor=blue
}
\usepackage[usenames, dvipsnames]{color}

\graphicspath{{.}{Figuras/}} 

\begin{document}

\title{Optimal control over the full counting statistics in a non-adiabatic pump}

\author{Fran\c{c}ois Impens} 
\affiliation{Instituto de F\'{\i}sica, Universidade Federal 
Rio de Janeiro, 21941-972 Rio de Janeiro, RJ, Brazil}
\author{Felippo M. D’Angelis}
\affiliation{Instituto de F\'{\i}sica, Universidade Federal 
Rio de Janeiro, 21941-972 Rio de Janeiro, RJ, Brazil}
\author{David Gu\'ery-Odelin}
\affiliation{Laboratoire Collisions Agr\'egats R\'eactivit\'e, UMR 5589, FERMI, UT3, Universit\'e de Toulouse, CNRS,
118 Route de Narbonne, 31062 Toulouse CEDEX 09, France}
\author{Felipe A. Pinheiro}
\affiliation{Instituto de F\'{\i}sica, Universidade Federal 
Rio de Janeiro, 21941-972 Rio de Janeiro, RJ, Brazil}
\author{Caio Lewenkopf}
\affiliation{Instituto de F\'{\i}sica, Universidade Federal 
Rio de Janeiro, 21941-972 Rio de Janeiro, RJ, Brazil}

\date{\today}
    
\begin{abstract}
We introduce a systematic procedure based on optimal control theory to address the full counting statistics of particle transport in a stochastic system.
Our approach enhances the performance of a Thouless pump in the non-adiabatic regime by simultaneously optimizing the average pumping rate while minimizing noise.
We demonstrate our optimization procedure on a paradigmatic model for the electronic transport through a quantum dot, both in the limit of vanishing Coulomb interaction and in the interacting regime.
Our method enables independent control of the moments associated with charge and spin transfer, allowing for the enhancement of spin current with minimal charge current or the independent tuning of spin and charge fluctuations. These results underscore the versatility of our approach, which can be applied to a broad class of stochastic systems.
\end{abstract}

\maketitle

\paragraph*{Introduction.--}
Thouless pumping \cite{Thouless1983, Citro2023thouless} captures the topological features of the quantum Hall effect enabling minimal dissipation charge and spin transfer in mesoscopic systems \cite{Switkes1999, Brouwer1998, Zhou1999, Mucciolo2002, Mares2004, Blumenthal2007, Hernandez2009, Croy2012}.
Its significance extends beyond condensed matter physics, sparking intense research interest across diverse scientific domains \cite{ Sinitsyn2007, Nakajima2016, Lohse2016, Oka2019, Fedorova2020observation,  Xia2021, Mostaan2022, Minguzzi2022,  Walter2023}. 
Notably, robust quantization of pumped charge has driven metrological advancements \cite{Pekola2013, Kaestner2015, Yamahata2016, Stein2017, Scherer2019, Hohls2022}.
At its core, pumping is realized by cyclically modulating the system-reservoir couplings, effectively pumping electrons at zero bias.
However, the delicate balance between pumping speed and adiabaticity presents a critical challenge. 
To minimize dissipation, the modulation must be sufficiently slow, ensuring the system's instantaneous probability vector closely tracks its stationary state. 
Operating beyond the adiabatic limit inevitably leads to deviations from the stationary state \cite{Shih1994}, resulting in a sharp decline in the geometric \cite{Xiao2010} pumping rate.

For a stochastic system coupled to multiple reservoirs, the stochastic dynamics equation bears a strong resemblance to a Schr\"odinger equation in imaginary time, where the system's instantaneous probability vector, $(p_1(t),...,p_N(t))$  associated with the different states $\{ 1,...,N \}$, describes the system state.
Within this framework, the breakdown of the unique features of a Thouless pumping at high frequencies can be understood as the emergence of undesired non-adiabatic transitions from the instantaneous adiabatic state.
To maintain the efficiency of geometric pumping in the non-adiabatic regime, previous studies~\cite{Nori2020, Takahashi2020} have applied \textit{shortcut-to-adiabaticity} (STA) techniques~\cite{Chen2010,Guery-Odelin2019,Guery-Odelin2023}, originally developed in quantum physics.
In practice, such methods often introduce an extra driving term— e.g. the counterdiabatic correction—to suppress non-adiabatic transitions.
However, this approach suffers from several limitations.
First, implementing the shortcut strategy is challenging when only a subset of the transition rates can be controlled.
Moreover, while STA can extend Thouless pumping beyond the adiabatic regime, the range of accessible frequencies remains limited: at sufficiently high frequencies, the required time-dependent transition rates to implement the counterdiabatic drive would become negative, rendering them unphysical.
Most importantly, although STA techniques successfully preserve the average number of particles transferred per cycle, they fail to control the fluctuations in the number of transferred particles.
This is a crucial limitation for many applications, particularly in metrology~\cite{Kaestner2008, Chorley2012, Bae2020}.
So far, fluctuations in geometric pumping have been mostly analyzed in the adiabatic limit~\cite{Potanina2019}.
Full counting statistics~(FCS) \cite{Levitov1996, Blanter2001, Nazarov2009} have been investigated in the non-adiabatic regime only for specific examples of Thouless cycles~\cite{Croy2016}, but these studies reported very low pumping efficiency at high frequencies.

Here, we present a systematic procedure for simultaneously controlling both the average flux and the counting statistics of the transfer.
To this end, we introduce an effective space of sufficiently large dimension to accommodate the system's state along with additional degrees of freedom governing the FCS dynamics. Applied to charge and spin transport through a quantum dot (QD), we show that our procedure demonstrates unprecedented independent control of charge and spin fluctuations, paving the way for time-dependent applications in spintronics.

\begin{figure}[t!]
\begin{center}
\includegraphics[width=6.5 cm]{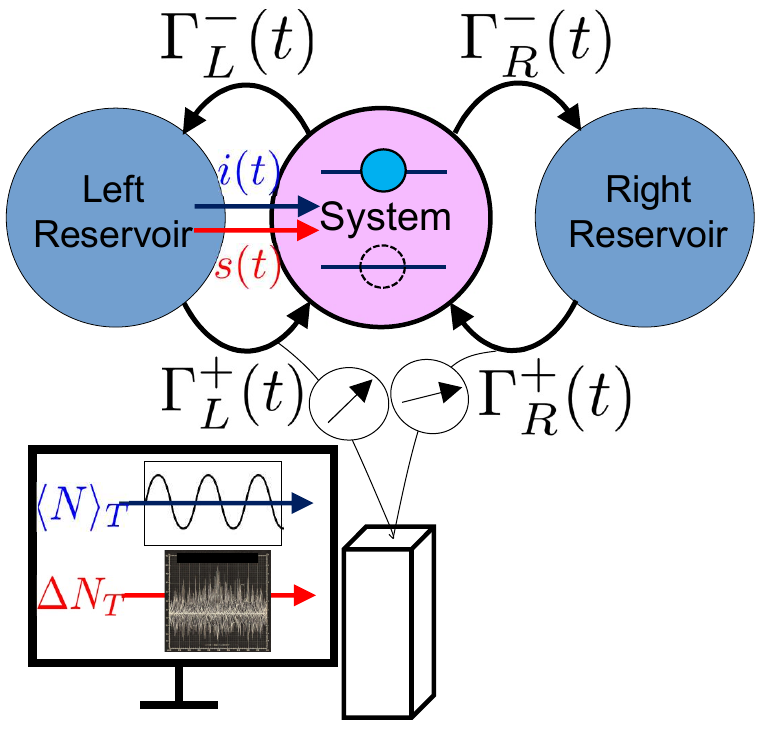}
\end{center}  
\caption{
Schematic model of a quantum dot coupled to two reservoirs. 
Our procedure optimizes $\Gamma_{L,R}^+(t)$ to maximize the average pumped current while minimizing the associated noise.
}
 \label{FigR}
\end{figure}

\paragraph*{FCS formalism for a two-state system.--}
We consider a stochastic system coupled to two reservoirs with time-dependent tunneling probabilities or decay widths
$\{\Gamma_{L}^{\pm}(t),\Gamma_{R}^{\pm}(t)\}$. 
$\Gamma_{L}^{+}(t)$ represents particle transfer from the left reservoir to the system, while $\Gamma_{L}^{-}(t)$ accounts for the reverse process, as shown in Fig.~\ref{FigR}. 
The system state is described by a probability vector $|\mathbf{p}(t) \rangle = (p_0(t), ..., p_N(t))^T$, where $p_k(t)$ denotes the probability of having $k$ particles in the system.
The coupling to the reservoirs induces a Markovian dynamics governed by the master equation
\begin{equation}
\partial_t | \mathbf{p}(t) \rangle =  \mathcal{L}_0(t)  | \mathbf{p}(t) \rangle.
\label{eq:MasterEquation}
\end{equation}
This stochastic equation captures the microscopic theory of quantum transport~\cite{Gurvitz96}. 
To illustrate our method, we consider a single-resonance QD coupled to two reservoirs, where the system can accommodate at most one particle~{(Fig.~\ref{FigR}). 
This simplification is valid when the coupling rates, $\Gamma_{L,R}^{\pm}(t)$ are significantly smaller than 
both the thermal energy and the chemical potential difference between the reservoirs, specifically
$\Gamma_{L,R}^{\pm}(t) \ll \max \{k_B T, eV \}$~\cite{Bagrets2003}.
The operator $\mathcal{L}_0(t)$, defined below (Eq.~\eqref{eq:Lchi} with $\chi=0$), depends on the transition rates $\Gamma_{L,R}^{\pm}(t)$. 
In accordance with the conservation of the total probability sum, $\langle \mathbf{q}_0| \mathbf{p}(t) \rangle = 1$, where $|\mathbf{q}_0 \rangle = (1,1)$, it follows that $\langle \mathbf{q}_0 | \mathcal{L}_0(t) = 0$ (each column of $\mathcal{L}_0(t)$ sums to zero). 
Consequently, $\mathcal{L}_0(t)$ has a zero eigenvalue associated with a stationary solution ${\bf \pi}(t) = (\pi_0(t), \pi_1(t))$, along with a negative eigenvalue $-\Gamma(t)$ where $\Gamma(t) =\Gamma_{L}^{+} +\Gamma_{L}^{-} +\Gamma_{R}^{+}+\Gamma_{R}^{-}$.

A FCS theory of charge transport~\cite{Levitov1996} can be formulated for this system~\cite{Bagrets2003}.
The characteristic function $\phi(\chi,t) = \langle e^{i \chi N} \rangle$ describes the transfer of $N$ particles from the system to the left reservoir over the time interval $[t_0,t]$.
Moments are obtained by differentiation with respect to the counting field $\chi$:
$\langle N^m \rangle = \left. \partial_{i \chi}^m \phi(\chi,t) \right|_{\chi=0}$, 
where $ \langle N^m \rangle \equiv \langle N^m \rangle(t)$.
To determine $\phi(\chi,t)$, 
we solve the master equation $\partial_t  |\mathbf{p}(\chi,t)\rangle  =  \mathcal{L}(\chi,t) |\mathbf{p}(\chi,t)\rangle$ with initial condition $|\mathbf{p}(\chi,t_0)\rangle=|\mathbf{p}(t_0)\rangle$ and a modified Lagrangian~\cite{Bagrets2003}:
\begin{equation}
 	\mathcal{L}(\chi,t) =\left( \begin{array}{cc}  -\Gamma_L^{+}(t)-\Gamma_R^{+}(t) & \Gamma_L^{-}(t) e^{-i \chi}+\Gamma_R^{-}(t) \\
 		\Gamma_L^{+}(t) e^{i \chi}+\Gamma_R^{+}(t) & -\Gamma_L^{+}(t)-\Gamma_R^{+}(t)\\
 	\end{array} \right) \label{eq:Lchi}
\end{equation}
 obtained from $\mathcal{L}_0(t)$ by modifying the left reservoir's tunneling rates as $\Gamma_{L}^{\pm}(t) \rightarrow \Gamma_L^{\pm}(t) e^{ \pm i \chi}$, by introducing a counting field factor $e^{\pm i \chi}$ in the off-diagonal elements. 
 This substitution affects only the left reservoir rates, as we monitor the particle current and 
the FCS there, and satisfies $\mathcal{L}_0(t)=\mathcal{L}(\chi=0,t)$.
The characteristic function is then $\phi(\chi,t) = \langle \mathbf{q}_0| \mathbf{p}(\chi,t) \rangle $. 
To obtain the average particle number and noise transfer in the steady state regime, we avoid transients by initiating the counting time $t_0$ when $|\mathbf{p}(t_0)\rangle$ is near the steady state solution of Eq.~\eqref{eq:MasterEquation}. Typically, $t_0=N_0 T$ with $N_0=3$ suffices for our numerical examples~(the system-reservoir couplings begin at $t=0$).}

 Let us now evaluate the steady state moments $\langle N^m \rangle_T =  \langle\left. \mathbf{q}_0 | \partial_{i \chi}^m  | \mathbf{p}(\chi,t_0+T)  \rangle \right|_{\chi=0} $ for the pumped particle number per period. 
 The first moment stems from the integration of the particle current $i(t)= \partial_t \langle N \rangle=  \langle  \mathbf{q}_0 |  \hat J_1(t) \: | \mathbf{p}(t)\rangle$ over a period, with  $\hat J_1(t)=  \left. \partial_{i \chi} \mathcal{L}(\chi,t) \right|_{\chi=0}$ the current operator~\cite{Croy2016}. The noise current is given by $s(t)=  \partial_t [ \langle N^2 \rangle - \langle N \rangle^2 ]= \langle \mathbf{q}_0 | \hat{J}_2(t) |  \mathbf{p} (t) \rangle + 2 \langle \mathbf{q}_0  | \hat{J}_1(t) | \overline{\mathbf{p}}(t) \rangle - 2  i(t)  \langle \mathbf{q}_0 | \overline{\mathbf{p}}(t)\rangle$, where we have introduced the operator $\hat{J}_2(t)=\left. \partial_{i \chi}^2 \mathcal{L}(\chi,t) \right|_{\chi=0}$ and the vector
$ | \overline{\mathbf{p}}(t) \rangle = \left. |\partial_{i \chi} \mathbf{p}(\chi,t) \rangle \right|_{\chi=0}$. 
The dynamical equation for $|\overline{\mathbf{p}}(t) \rangle $ is readily obtained by differentiating the master equation with respect to the counting field $\chi$, namely, $\partial_t |\overline{\mathbf{p}}(t) \rangle =  \mathcal{L}_0(t) |\overline{\mathbf{p}}(t)\rangle +  \hat{J}_1(t) |\mathbf{p}(t) \rangle$. 
The initial condition is $|\overline{\mathbf{p}}(t_0) \rangle=(0,0)^T$.
\paragraph*{Shortcut-to-adiabaticity and the FCS.--}
Here, we assess how a standard STA technique~\cite{Guery-Odelin2019}, recently applied to nonadiabatic Thouless pumping~\cite{Nori2020, Takahashi2020}, affects the FCS. 
In the STA approach, an additional counterdiabatic (CD) driving term forces the instantaneous probability vector $| \mathbf{p}(t) \rangle$ to remain close to the adiabatic trajectory throughout the cycle.
The associated CD driving constrains only the total transition rates, $\Gamma^{\pm}(t) = \Gamma^{\pm}_L(t) + \Gamma^{\pm}_R(t)$, without imposing specific values for the individual left and right rates. This introduces a useful gauge freedom in the shortcut implementation.

As an example, we consider $\Gamma_L^{0 (+)}(t)=A+R \cos (\Omega t),$ $\Gamma_R^{0 (+)}(t)=A+R \sin (\Omega t),$ and $\Gamma_{L,R}^{0 (-)}(t)=1$~\cite{Nori2020}. 
By implementing the appropriate CD shortcut on the right side, we get: $\Gamma_L^{{\rm STA}(\pm)}(t) = \Gamma_L^{0 (\pm)}(t)$ and $\Gamma_R^{{\rm STA}(\pm)}(t)= \Gamma_R^{0 (\pm)}(t) \pm \gamma(t)$.
Here, the extra CD term reads $\gamma(t)= \partial_t \left[ \partial_t \left( \Gamma^{0-}(t) / \Gamma^0(t) \right) / \Gamma^0(t) \right]$ and $\Gamma^0(t)=\Gamma^{0+}(t)+\Gamma^{0-}(t)$. 

We analyze the FCS for the non-adiabatic cycling frequency $\Omega = 10 \Gamma_0$. 
While the STA driving achieves near-perfect average particle transfer, with $\langle N \rangle_{T,\rm STA} / N_{T, \rm geo} \simeq 1$, approaching the maximum geometric value $N_{T,\rm geo} = 2 \pi R^2 / (4 (A+1)^2 - 2 R^2)^{3/4} = 6.5 \times 10^{-3}$ obtained in the adiabatic limit, and significantly outperforming of the original protocol ($\langle N \rangle_{T,\rm (0)} \simeq 0.5 N_{T,\rm geo}$), it has minimal impact on the noise.
The noise remains much larger than the geometric pumping rate for both protocols: $(\Delta N)_{T, \rm STA}^2 \simeq (\Delta N)_{T, \rm (0)}^2 \simeq 0.5$,  which is detrimental to many applications.
To address this, a more sophisticated approach is required, considerering the dynamics of both $| \mathbf{p} \rangle$ and  $| \overline{\mathbf{p}} \rangle$, as detailed below.

\paragraph*{Optimal effective FCS quantum control.--}
The dynamics for the global vector $| \mathbf{p}, \overline{\mathbf{p}} \rangle =( \langle\mathbf{p} |, \langle \overline{\mathbf{p}}|)^T$ is compactly expressed as 
$\partial_t | \mathbf{p}, \overline{\mathbf{p}} \rangle=\mathcal{L}_4(\mathbf{u}(t), t) | \mathbf{p},\overline{\mathbf{p}} \rangle$, where  
\begin{equation}
\mathcal{L}_4(\mathbf{u}(t), t)  =   \left( \begin{array}{cc} \mathcal{L}_0(t) & 0  \\  \hat{J}_1(t) & \mathcal{L}_0(t)  \end{array} \right),
\label{eq:forwardpropagation1}
\end{equation}
and $\mathbf{u}(t)=(\Gamma_L^+(t),\Gamma_R^+(t),\Gamma_L^-(t),\Gamma_R^-(t))$ is the control vector. 

The goal is to maximize particle transfer while minimizing noise in the cyclic evolution of the control parameter.
This conditional optimization problem can be readily formulated using the Pontryagin Maximum Principle (PMP)~\cite{PontryaginBook}. 
For a system described by the differential equations $\dot{\mathbf{x}}= f(\mathbf{x}(t),\mathbf{u}(t),t)$, the PMP provides a framework to find the control vector $\mathbf{u}(t)$ that minimizes the cost function $C[\mathbf{u}(t)]= \int_0^T dt f_0(\mathbf{x}(t),\mathbf{u}(t),t)$.
This is achieved by an effective Hamiltonian formalism, where the  Pontryagin Hamiltonian is defined as $H_P (\mathbf{x}(t),\mathbf{v}(t),\mathbf{u}(t),t) \! \! = \! \!  -f_0(\mathbf{x}(t),\mathbf{u}(t), t) \! + \! \:^{T} \mathbf{v}(t) \cdot f(\mathbf{x}(t),\mathbf{u}(t),t)$.
Here, $\mathbf{v}(t)$ is an adjoint vector (a generalized Lagrange multiplier). 
The optimal solution $(\mathbf{x}^*(t),\mathbf{v}^*(t),\mathbf{u}^*(t))$ for an unbound control parameter is obtained by solving $\dot{\mathbf{v}}= - \partial_{\mathbf{x}} H_P$, $\dot{\mathbf{x}}= \partial_{\mathbf{v}} H_P$ and $\partial_{\mathbf{u}} H_P|_{\mathbf{u}^*} =0$. 
In our case, $f \equiv \mathcal{L}_4$ and $f_0= w_1 i+ w_2 s$, with $i(t) = \Gamma_L^+ x_1 -\Gamma_L^- x_2$ and $s(t)= \Gamma_L^+ x_1+\Gamma_L^- x_2 + 2 \Gamma_L^+ x_3-2\Gamma_L^- x_4-2 (\Gamma_L^+ x_1 -\Gamma_L^- x_2)(x_3+x_4)$. 
To minimize fluctuations, we set $w_2>0$.
The sign of $w_1$ determines the direction of the particle current: $w_1<0$ for currents flowing {\sl from} the left reservoir $i(t)>0$, and $w_1>0$ for currents {\sl towards} it, $i(t)<0$.
Since the trajectory endpoint $\mathbf{x}(T)$ is kept free, the boundary condition for the adjoint is $\mathbf{v}(T) = \mathbf{0}$. 
These optimal solutions are obtained numerically using an iterative gradient descent optimization routine on a discrete time grid \cite{Sugny21,Ansel_2024}. 

\begin{figure}[t!]
\begin{center}
\includegraphics[width=8.5cm]{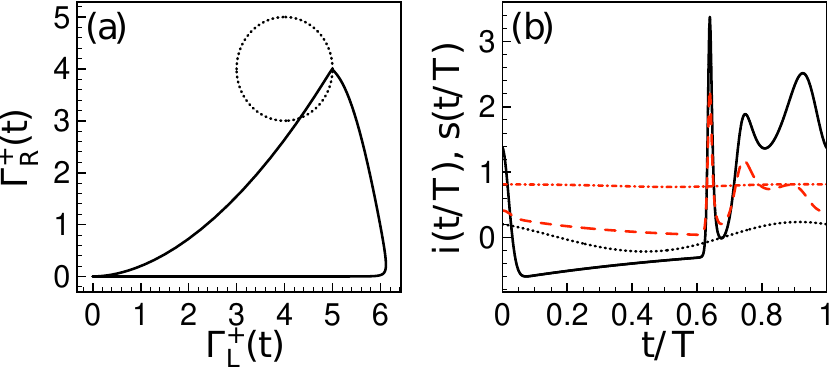}
\end{center}  \caption{Optimal transfer: 
(a) Parametric plot of $ \Gamma_L^+(t)$ versus $\Gamma_R^+(t)$ showing the initial (dashed line) and optimized (solid line) pumping rate cycle. 
(b) Final particle $i(t/T)$~(solid line) and noise $s(t/T)$~(red dashed line) currents, and their initial profiles~(dotted line for $i(t/T)$ and red dash-dotted line for $s(t/T)$).
All results are plotted as a function of the dimensionless time $t/T$ with $T=2\pi/\Omega$ the cycle period. 
These results obtained after 100 iterations for $w_1=-0.2=-w_2$, $\epsilon(t)=0.8 \Gamma_0 \sin(\pi t/T)$, and $\Omega=10 \Gamma_0$.}
 \label{fig:PlotOptimisation1}
\end{figure}

\paragraph*{Optimal control in a two-state system.--}
We present a practical implementation of this method for a two-state stochastic system (representing an empty or occupied single particle 
state), with FCS described by the Lagrangian ~\eqref{eq:Lchi}. 
This system corresponds to a QD in the limit of vanishing Coulomb interaction, allowing for independent treatment of spin-up and spin-down electron transport.
We begin with the time-dependent pumping rates  $\Gamma_{L,R}^{0 \pm}(t)$, 
as previously defined. 
To maintain the periodicity and positivity of the pumping rates during the iterative process, we define $\Gamma_{L,R}^{+}(t)=[\sqrt{\Gamma_L^{0 +}(t)}+ {\sin }(\Omega t /2)\, f_{L,R}(t)]^2$, and allow the functions $f_{L,R}(t)$  to evolve freely according to the optimal control routine, while leaving the rates $\Gamma_{L,R}^{-}(t)$ unchanged. 
Note that the equation governing the adjoint vector $\mathbf{v}(t)$, $\dot{\mathbf{v}}=-\partial_{\mathbf{x}}f_0-\mathcal{L}_4^T\cdot\mathbf{v}$, differs from the one for the propagation of the system state $\mathbf{x}$.
Since the operator $\mathcal{L}_4$ governing the forward propagation has negative (and null) eigenvalues, the operator $- \mathcal{L}_4^T$ appearing in the adjoint dynamics typically has positive (or null) eigenvalues. 
This prevents exponential growth during the backward propagation.
Figure~\ref{fig:PlotOptimisation1} shows the results of our optimization procedure. 
At a frequency $\Omega=10 \Gamma_0$, we achieve $\langle N \rangle_T = 0.22$ and $\Delta N_T^2 = 0.23$. 
Notably, $\Delta N_T^2$ is reduced by approximately $50 \%$ compared to the shortcut trajectory, while $\langle N \rangle_T$ is increased by over an order of magnitude.

\paragraph*{Optimal control in the 
interacting regime.--}
To demonstrate the versatility of our method, we now explore the 
interacting regime, 
electrons are pumped into the system at different rates depending on their spin orientation. 
Let us consider a single-resonance QD at the Coulomb blockade regime in the limit of large charging energy
~\cite{Bagrets2003}. 
In this regime, the system can be either empty, occupied by a single spin-up or spin-down electron.
The corresponding FCS Lagrangian, expressed in the $\{ | 0 \rangle, | \uparrow \rangle, | \downarrow \rangle \}$ basis, is given by:
\begin{widetext}
\begin{equation}
\label{eq:LagrangianL3}
\mathcal{L}(\boldsymbol{\chi},t)= \left( \begin{array}{ccc} -\Gamma^{+}(t) &  \Gamma_{L,\uparrow}^{-}(t) e^{-i \chi_{\uparrow}} +  \Gamma_{R,\uparrow}^{-}(t)    & \Gamma_{L,\downarrow}^{-}(t) e^{-i \chi_{\downarrow}}  +  \Gamma_{R,\downarrow}^{-}(t) \\
\Gamma_{L,\uparrow}^{+}(t) e^{i \chi_{\uparrow}} +  \Gamma_{R,\uparrow}^{+}(t) & -\Gamma_{\uparrow}^{-}(t) & 0   \\
\Gamma_{L,\downarrow}^{+}(t) e^{i \chi_{\downarrow}} +  \Gamma_{R,\downarrow,}^{+}(t) & 0 & -\Gamma_{\downarrow}^{-}(t) 
\end{array} \right).\,
\end{equation}
\end{widetext}
where $\Gamma^{+}(t)= \sum_{k=\{L,R\}} \sum_{\sigma=\{\uparrow,\downarrow\}} \Gamma_{k,\sigma}^{+}(t)$ and 
$\Gamma_{\uparrow}^{-}(t)=\Gamma_{L,\uparrow}^{-}(t)+\Gamma_{R,\uparrow}^{-}(t)$.  
We denote the average number of spin-up (spin-down) particles transferred from the left reservoir to the system per cycle as  $\langle N_{\uparrow} \rangle_T$ ($\langle N_{\downarrow} \rangle_T$), with corresponding current $i_{\uparrow}(t)$ ($i_{\downarrow}(t)$).
We associate distinct counting fields, $\chi_{\uparrow}$ and $\chi_{\downarrow}$, with  $i_{\uparrow}(t)$ and $i_{\downarrow}(t)$. 
The probability vector $|\mathbf{p}(t) \rangle$ follows Eq.~\eqref{eq:MasterEquation} with $\mathcal{L}_0(t)=\mathcal{L}(\boldsymbol{\chi}=\mathbf{0},t)$.

We also consider a more realistic model where the transition rates $\Gamma_{ k,\sigma}^{\pm}(t)$ are not independent but are determined by experimentally accessible physical parameters, such as applied voltages and magnetic fields, which influence the reservoir filling factors. 
The sharp dependence of these filling factors on the modulated fields leads to abrupt variations in the transition rates $\Gamma_{ k,\sigma}^{\pm}(t)$.
Consequently, the CD-STA approach becomes invalid, as the required counterdiabatic driving term involves extremely large or even unphysical negative $\Gamma$'s. 
In contrast, our PMP-based optimization procedure remains applicable to these available physical parameters.

We consider the QD model used in Refs.~\cite{Bagrets2003,Croy2016}, which describes spin transport in the Coulomb-blockade regime. 
The transition rates are given by $\Gamma_{k \sigma}^{(+)}(t)=f_{k \sigma}(t) \Gamma_k(t)$ and $\Gamma_{k \sigma}^{(-)}(t)=[1-f_{k \sigma}(t)] \Gamma_k(t)$, where $k \in \{ L,R \}$ denotes the reservoir, $\sigma \in \{ \uparrow, \downarrow \}$ the electron spin projection, and  $f_{k \sigma}(t)=[1+e^{- \epsilon_{k \sigma}(t)/(k_{B} T_{\rm sys})}]^{-1}$, the filling factors. 
The chemical potential $\epsilon_{k \sigma}(t)$ can be tuned by voltage gates and magnetic fields. 
Specifically, $\epsilon_{k \sigma}(t)=V_{k}(t) +(-1)^s \epsilon_{k Z}(t)$ consists of an electric voltage contribution $V_{k}(t)$ (identical for both spins) and a spin-dependent Zeeman contribution $(-1)^s \epsilon_{k Z }(t)$ induced by a tunable magnetic field (where $\sigma=+1$ corresponds to spin-up). 

\begin{figure}[t!]
\begin{center}
\includegraphics[width=8cm]{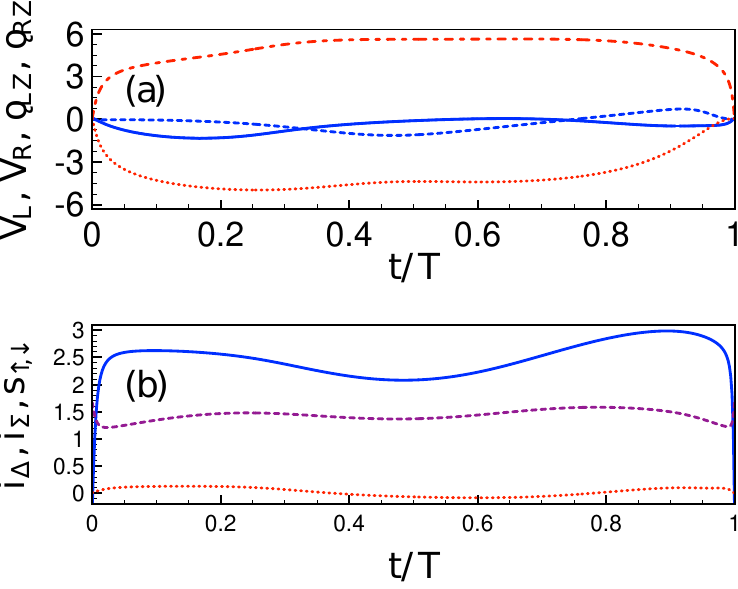}
\end{center}  \caption{Optimal transfer in a spin system after $100$ iterations: (a) Optimized control parameters $V_L(t/T)$~(solid blue line), $V_R(t/T)$~(dashed blue line), $\epsilon_{L Z}(t/T)$~(dash-dotted red line), $\epsilon_{R Z}(t/T)$~(dotted red line) in units of $V_0=k_B T_{\rm sys}$ as a function of $t/T$. 
(b) Dimensionless currents $i_{\Delta}(t/T)$~(solid blue line), $i_{\Sigma}(t/T)$~(dotted purple line) and spin fluctuations $s_{\uparrow,\downarrow}(t/T)$~(dashed red line) for the optimized cycle as a function of $t/T$. 
Parameters: $\Omega=10 \Gamma_0,$ $w_1=-0.5$, $w_2=1,$ $w_3=0.2$, numerical increment $\epsilon(t)=2 V_0 \sin(\pi t/T)$ with $V_0=k_B T_{\rm sys}$. 
Prescribed cycle: $\Gamma_L(t)=A+R \cos(\Omega t)$ and $\Gamma_R(t)=A+R \sin(\Omega t)$ with $A=4 \Gamma_0,$ $R= \Gamma_0$.  
Initial control parameters: $V_{L}(t)= (V_0/10) \cos (\Omega t)$, $V_{R}(t)= (V_0/10)  \sin (\Omega t)$ and $\epsilon_{LZ}(t)=  \epsilon_{RZ}(t) = V_0/20$.} 
 \label{fig:PlotOptimisationSpins}
\end{figure}

Next, we investigate the feasibility of generating a net average spin current while minimizing
fluctuations and total charge transport, which has potential applications in
spintronics~\cite{Fert}. 
The charge and spin currents are given by $i_q(t)= -e i_{\Sigma}(t)$ with $i_{\Sigma}(t)= i_{\uparrow}(t)+i_{\downarrow}(t)$, and $i_\sigma(t)=\frac {\hbar} {2}  i_{\Delta}(t)$ where $i_{\Delta}(t) = i_{\uparrow}(t)-i_{\downarrow}(t)$. 
The total charge and spin transfer per cycle are proportional to the dimensionless quantities
$\langle N \rangle_T = \langle N_{\uparrow}+N_{\downarrow} \rangle_T = \int_{t_0}^{t_0+T} i_{\Sigma}(t) dt$ and $\langle \tilde{S} \rangle_T = \langle N_{\uparrow}-N_{\downarrow} \rangle_T =\int_{t_0}^{t_0+T} i_{\Delta}(t) dt $. 
Similarly, we quantify the dimensionless spin fluctuations per cycle as 
$\Delta \tilde{S}_T= \int_{t_0}^{t_0+T} s_{\uparrow \downarrow}(t) dt$ with $s_{\uparrow \downarrow}(t)=\partial_t \left[ \langle (N_{\uparrow} -N_{\downarrow})^2 \rangle - \langle N_{\uparrow} -N_{\downarrow} \rangle^2 \right]$, 
where higher moments are given by  $\langle N_{\uparrow }^{m_1}  N_{\downarrow }^{m_2}  \rangle =  \langle \mathbf{q}_0 |  \partial_{i \chi_{\uparrow}}^{m_1} \partial_{i \chi_{\downarrow}}^{m_2} | \mathbf{p}(\boldsymbol{\chi},t) \rangle |_{\boldsymbol{\chi}=\mathbf{0}}$ with $|\mathbf{q}_0 \rangle=(1,1,1)^T$  ($\langle \mathbf{q}_0 | \mathcal{L}_0(t)=0$). 

The spin currents read $i_\sigma(t) = \langle \mathbf{q}_0 | \hat{J}_{1,\sigma}| \mathbf{p}(t) \rangle$ with $\hat{J}_{1,\sigma}= \Gamma_{L, \sigma}^{(+)}(t)|\sigma \rangle \langle 0| - \Gamma_{L, \sigma}^{(-)}(t)  |0\rangle \langle \sigma|$. 
In turn, the spin fluctuations read
$s_{\uparrow \downarrow}(t)= \langle \mathbf{q}_0 | \mathbf{S}_{\uparrow \downarrow}(t) \rangle$ 
with \cite{SM}
\begin{align*}
    \mathbf{S}_{\uparrow \downarrow}(t)&\, =  \left[\hat{J}_{2,\uparrow} (t)+ \hat{J}_{2,\downarrow} (t) \right ] | \mathbf{p}(t)\rangle + 
    2 \sum_\sigma \hat{J}_{1,\sigma}  (t)  |\overline{\mathbf{p}}^\sigma(t)\rangle \\& 
    - 2 \sum_\sigma \hat{J}_{1,\sigma} (t)|\overline{\mathbf{p}}^{\bar{\sigma}}(t)\rangle  
    - 2 i_{\Delta}(t) \left[|\overline{\mathbf{p}}^{\uparrow}(t)\rangle-|\overline{\mathbf{p}}^\downarrow(t)\rangle \right],
\end{align*}
where $\bar\sigma=-\sigma$. Here, 
$|\overline{\mathbf{p}}^\sigma\!(t)\rangle=\partial_{i \chi_\sigma} |\mathbf{p}(\boldsymbol{\chi},t)\rangle |_{\boldsymbol{\chi}=0}$ 
evolve dynamically under the Lagrangian $\mathcal{L}_0(t)$, driven by their coupling to the probability vector $|\mathbf{p}(t)\rangle$ through their respective current operators $\hat{J}_{1,\sigma}(t)$ and fulfill the initial condition $|\overline{\mathbf{p}}^\sigma(t_0)\rangle=(0,0,0)^T$.
In summary, the FCS dynamics of this system is governed by a master equation in a 9-dimensional space, $\partial_t | \mathbf{p}, \overline{\mathbf{p}}^{\uparrow},\overline{\mathbf{p}}^{\downarrow} \rangle=      \mathcal{L}_9(\mathbf{u}(t), t)  | \mathbf{p}, \overline{\mathbf{p}}^{\uparrow},\overline{\mathbf{p}}^{\downarrow} \rangle$, where   
 \begin{eqnarray}
 \mathcal{L}_9(\mathbf{u}(t), t)  =   \left( \begin{array}{ccc} \mathcal{L}_0(t) & 0  & 0 \\  \hat{J}_{1,\uparrow}(t) & \mathcal{L}_0(t)  & 0 \\ \hat{J}_{1,\downarrow}(t) & 0  & \mathcal{L}_0(t) \end{array} \right) .
\label{eq:forwardpropagation2}
\end{eqnarray}
We define the cost function as $f_0(t)= w_1 i_{\Delta}(t)+ w_2 i_{\Sigma}(t)^2+ w_3  s_{\uparrow \downarrow}(t),$
where the positive weights $w_{2,3}>0$ are chosen to suppress the average charge current and reduce fluctuations in spin transfer. 
A negative $w_1<0$ ($w_1>0$) favors a net rightward transfer of spin-up (spin-down) particles.

We consider the cyclic couplings $\Gamma_L(t)=A+R \cos(\Omega t)$ and $\Gamma_R(t)=A+R \sin(\Omega t)$, and optimize the parameters $\{ V_{L}(t),V_{R}(t),\epsilon_{LZ}(t),\epsilon_{RZ}(t) \}$. 
To enforce the periodicity of the parameters, we apply a time-dependent increment $\epsilon(t)= \epsilon_0 \sin(\pi t/T)$ at each iteration, ensuring that initial and final values remain unchanged.
Figure~\ref{fig:PlotOptimisationSpins} displays the resulting optimized cycle and the corresponding time-dependent currents $i_{\Delta}(t)$, $i_{\Sigma}(t)$, and spin fluctuations $s_{\uparrow \downarrow}(t)$. 
The optimization process yields a nearly pure spin current~($\langle \tilde{S} \rangle_{T, \rm opt} = 1.55$, $\langle N \rangle_{T, \rm opt} =0.02$) with the dimensionless spin fluctuations $\Delta \tilde{S}^2_{T, \rm opt} =0.90$, in contrast to a negligible 
initial spin current $\langle \tilde{S} \rangle_{T, \rm ini}= 1.4 \times 10^{-3}$ and  $\Delta \tilde{S}^2_{T, \rm ini}  = 0.83$.
The signal-to-noise ratio is improved by several orders of magnitude, while effectively suppressing the charge current.
We note that our method also allows for the decoupling of spin and charge transfer statistics. 
In this case, the cost functional involves simultaneously the spin $s_{\uparrow \downarrow}(t)$ and particle $s_{N}(t)$ fluctuations~\cite{SM}.

\paragraph*{Conclusions.--} 
We have investigated a nonadiabatic electron pump using full counting statistics with an approach that optimizes both the current and its noise. We demonstrate that our approach matches the STA's efficiency in generating a high-fidelity quantized current, but significantly expands its capabilities through a wider frequency range and the ability to mitigate current fluctuations. We have applied our method to demonstrate optimal control over a Thouless pump in the non-adiabatic regime in which charge and spin fluctuation can be independently tuned. This method is particularly relevant for high-accuracy pumping and can be readily extended to control even higher-order moments, $\langle N^m \rangle_T$ with $m \geq 3$~\cite{ShortcutSynchro23}. 
As such, it can be transposed to a wide range of quantum phenomena in mesoscopic physics \cite{PhysRevE.57.7297,doi:10.1073/pnas.0708040104,PhysRevLett.101.140602,PhysRevE.84.051110,PhysRevB.86.235308}, including charge transport in nanostructures \cite{BRANDES2005315}, stochastic thermodynamics \cite{PhysRevLett.99.180601,RevModPhys.81.1665,Seifert_2012,Bonanca14,PhysRevE.96.022118,PhysRevE.98.010104}, heat transfer \cite{RevModPhys.83.131}, and Brownian motors \cite{RevModPhys.81.387}.

\begin{acknowledgments}
    This work was supported by the Brazilian funding agencies CNPq, CAPES, and FAPERJ. 
\end{acknowledgments}

\bibliography{shortcuts}

\newpage

\begin{widetext}
\section{Supplemental Material}

This Supplemental Material provides technical details involved in the derivation of charge/spin intensity and noise currents for the spin transport in the interacting regime. It also presents additional information on the reduction of
correlations between spin and charge counting statistics.\\

\section*{S1: FCS in the Spin Transport}
\label{SM:FCS-spin-transport }

Our starting point is the FCS Lagrangian for the transfer of spin $1/2$-particles in the strongly interacting regime~[Eq.(4) of the main text]
\begin{equation}
\mathcal{L}(\boldsymbol{\chi},t)= \left( \begin{array}{ccc} -\Gamma^{(+)}(t) &  \Gamma_{L,\uparrow}^{(-)}(t) e^{-i \chi_{\uparrow}} +  \Gamma_{R,\uparrow}^{(-)}(t)    & \Gamma_{L,\downarrow}^{(-)}(t) e^{-i \chi_{\downarrow}}  +  \Gamma_{R,\downarrow}^{(-)}(t) \\
\Gamma_{L,\uparrow}^{(+)}(t) e^{i \chi_{\uparrow}} +  \Gamma_{R,\uparrow}^{(+)}(t) & -\Gamma_{\uparrow}^{(-)}(t) & 0   \\
\Gamma_{L,\downarrow}^{(+)}(t) e^{i \chi_{\downarrow}} +  \Gamma_{R,\downarrow}^{(+)}(t) & 0 & -\Gamma_{\downarrow}^{(-)}(t) 
\end{array} \right)
\end{equation}
with $\boldsymbol{\chi}=(\chi_{\uparrow},\chi_{\downarrow}).$
The successive moments are given by $\langle N_{\uparrow }^{m_1}  N_{\downarrow }^{m_2}  \rangle = \langle \mathbf{q}_0 |  \partial_{i \chi_{\uparrow}}^{m_1} \partial_{i \chi_{\downarrow}}^{m_2} | \mathbf{p}(\boldsymbol{\chi},t) \rangle  |_{\boldsymbol{\chi}=\mathbf{0}} $ with $| \mathbf{q}_0 \rangle =(1,1,1)^T$ s.t. $\langle \mathbf{q}_0 | \mathcal{L}(\mathbf{0},t)=0$ at all times. Consistently with the Coulomb-blockade enabling at most one particle in the system, one retrieves $\langle N_{\uparrow }^{m_1}  N_{\downarrow }^{m_2}  \rangle=0$ if $m_1 m_2 \neq 0$. The current associated to spin-up particles is obtained as 
\begin{eqnarray}
i_{\uparrow}(t) & = & \langle \mathbf{q}_0 | \left. \partial_t \partial_{i \chi_{\uparrow}} \mathbf{p}(\boldsymbol{\chi},t) \rangle  \right|_{\boldsymbol{\chi}=\mathbf{0}}  = \langle \mathbf{q}_0 | \hat{J}_{1,\uparrow}  | \mathbf{p}(t) \rangle
\end{eqnarray}
We introduce the operators
$$
\hat{J}_{1,\uparrow} =\partial_{i \chi_{\uparrow}} \mathcal{L}(\boldsymbol{\chi},t) |_{\boldsymbol{\chi}=0}  = \left( \begin{array}{ccc} 0 & -\Gamma_{L,\uparrow}^{(-)}(t)    & 0 \\
\Gamma_{L,\uparrow}^{(+)}(t) & 0 & 0   \\
0 & 0 & 0
\end{array} \right)  \quad \mbox{and} \quad 
\hat{J}_{1,\downarrow} =\partial_{i \chi_{\downarrow}} \mathcal{L}(\boldsymbol{\chi},t) |_{\boldsymbol{\chi}=0}  = \left( \begin{array}{ccc} 0 & 0 &  -\Gamma_{L,\downarrow}^{(-)}(t)     \\
0 & 0 & 0   \\
\Gamma_{L,\downarrow}^{(+)}(t)  & 0 & 0
\end{array} \right) 
$$
so that $i_{\uparrow}(t) =  \Gamma_{L,\uparrow}^{(+)} p_0 - \Gamma_{L,\uparrow}^{(-)} p_{\uparrow}$ and $i_{\downarrow}(t) =  \Gamma_{L,\downarrow}^{(+)}(t) p_0 - \Gamma_{L,\downarrow}^{(-)}(t) p_{\downarrow}$. 
The spin transferred per cycle corresponds to $\langle \tilde{S} \rangle_T = \frac {\hbar} {2}\langle N_{\uparrow} - N_{\downarrow} \rangle_T.$ The charge current and the spin current read respectively $i_q(t)=q i_{\Sigma}(t)$ with $i_{\Sigma}(t)= i_{\uparrow}(t)+i_{\downarrow}(t)$
and $q=-e$, and $i_{\sigma}(t)=\frac {\hbar} {2}  i_{\Delta}(t)$ with $i_{\Delta}(t) = i_{\uparrow}(t)-i_{\downarrow}(t)$. 
The spin fluctuations are calculated following the procedure as $\Delta \tilde{S}_T^2= \frac {\hbar^2} {4} \int_{t_0}^{t_0+T} s_{\uparrow \downarrow}(t) dt$ with
\begin{eqnarray}
s_{\uparrow \downarrow}(t) &  = &  \partial_t \left[ \langle (N_{\uparrow} -N_{\downarrow})^2 \rangle - \langle N_{\uparrow} -N_{\downarrow} \rangle^2 \right] \nonumber \\
 &  = &  \langle \mathbf{q}_0 | \left( \partial_{i \chi_{\uparrow}}-\partial_{i \chi_{\downarrow}} \right)^2 \left[ \frac {} {} \mathcal{L}(\boldsymbol{\chi},t) | \mathbf{p}(\boldsymbol{\chi},t) \rangle \right] \:  |_{\boldsymbol{\chi}=\mathbf{0}} -  2 i_{\Delta}(t) \langle \mathbf{q}_0 | \left( | \overline{\mathbf{p}(t)}^{\uparrow} \rangle- | \overline{\mathbf{p}(t)}^{\downarrow} \rangle \right)  \nonumber \\
  &  = &   \langle \mathbf{q}_0 | \left(\hat{J}_{2,\uparrow} (t)+ \hat{J}_{2,\downarrow} (t) \right) | \mathbf{p}(t) \rangle + 2 \langle \mathbf{q}_0 | \hat{J}_{1,\uparrow}  (t) |  \overline{\mathbf{p}(t)}^{\uparrow} \rangle+ 2 \langle \mathbf{q}_0 | \hat{J}_{1,\downarrow} (t) | \overline{\mathbf{p}(t)}^{\downarrow} \rangle \nonumber \\
& - &    2 \langle \mathbf{q}_0 | \hat{J}_{1,\uparrow}  (t) | \overline{\mathbf{p}(t)}^{\downarrow} \rangle -  2 \langle \mathbf{q}_0 | \hat{J}_{1,\downarrow} (t) | \overline{\mathbf{p}(t)}^{\uparrow} \rangle -  2 i_{\Delta}(t) \langle \mathbf{q}_0 |\left(|\overline{\mathbf{p}(t)}^{\uparrow} \rangle- |\overline{\mathbf{p}(t)}^{\downarrow} \rangle \right)  
\end{eqnarray}
where we have used that $\partial_{i \chi_{\uparrow}} \partial_{i \chi_{\downarrow}}  \mathcal{L}(\boldsymbol{\chi},t) |_{\boldsymbol{\chi}=0}=0$ and $\langle \mathbf{q}_0 |  \mathcal{L}(\mathbf{0},t)=0$ between the two and the third lines. One observes the presence of crossed contributions that simultaneously involve both spin orientations. We have introduced the operator and the three-dimensional vector
\begin{eqnarray}
\hat{J}_{2,\uparrow}(t) =   \partial_{i \chi_{\uparrow}}^2  \mathcal{L}(\boldsymbol{\chi},t) |_{\boldsymbol{\chi}=0} = \left( \begin{array}{ccc} 0 & \Gamma_{L,\uparrow}^{(-)}(t)    & 0 \\
\Gamma_{L,\uparrow}^{(+)}(t) & 0 & 0   \\
0 & 0 & 0
\end{array} \right) \qquad | \overline{\mathbf{p}(t)}^{\uparrow} \rangle= \partial_{i \chi_{\uparrow}} | \mathbf{p}(\boldsymbol{\chi},t) \rangle |_{\boldsymbol{\chi}=0} 
\end{eqnarray}
$\hat{J}_{2,\downarrow}(t)$ and $|\overline{\mathbf{p}(t)}^{\downarrow} \rangle$ are defined analogously. $|\overline{\mathbf{p}(t)}^{\uparrow (\downarrow)} \rangle$ follows the differential equation $\partial_t |\overline{\mathbf{p}(t)}^{\uparrow (\downarrow)} \rangle = \mathcal{L}(\mathbf{0},t) | \overline{\mathbf{p}(t)}^{\uparrow (\downarrow)} \rangle+\hat{J}_{1,\uparrow (\downarrow)} (t) | \mathbf{p}(t) \rangle$. Summing up, one has now the coupled dynamics in a $3 \times 3$ dimensional space:
\begin{eqnarray}
 & & \frac {\partial | \mathbf{p}, \overline{\mathbf{p}}^{\uparrow},\overline{\mathbf{p}}^{\downarrow} \rangle} {\partial t}    =      \mathcal{L}_9(\mathbf{u}(t), t)  | \mathbf{p}, \overline{\mathbf{p}}^{\uparrow},\overline{\mathbf{p}}^{\downarrow} \rangle  \\ 
 & & | \mathbf{p}, \overline{\mathbf{p}}^{\uparrow},\overline{\mathbf{p}}^{\downarrow} \rangle =\left( \begin{array}{c}  \mathbf{p} \\  \overline{\mathbf{p}}^{\uparrow} \\  \overline{\mathbf{p}}^{\downarrow}  \end{array} \right) \qquad  \mathcal{L}_9(\mathbf{u}(t), t)  =   \left( \begin{array}{ccc} \mathcal{L}_0(t) & 0  & 0 \\  \hat{J}_{1,\uparrow}(t) & \mathcal{L}_0(t)  & 0 \\ \hat{J}_{1,\downarrow}(t) & 0  & \mathcal{L}_0(t) \end{array} \right) \nonumber
\label{eq:forwardpropagation1}
\end{eqnarray}

One finds then
\begin{eqnarray}
i_{\Delta}(t)=(\Gamma_{L,\uparrow}^{(+)} - \Gamma_{L,\downarrow}^{(+)}) p_0 - \Gamma_{L,\uparrow}^{(-)} p_{\uparrow}+ \Gamma_{L,\downarrow}^{(-)} p_{\downarrow} \nonumber \\
i_{\Sigma}(t)=(\Gamma_{L,\uparrow}^{(+)} + \Gamma_{L,\downarrow}^{(+)}) p_0 - \Gamma_{L,\uparrow}^{(-)} p_{\uparrow}- \Gamma_{L,\downarrow}^{(-)} p_{\downarrow} 
\end{eqnarray}
and
\begin{eqnarray}
 s_{\uparrow \downarrow}(t) &  = & \Gamma_{L,\uparrow}^{(+)} p_0 +\Gamma_{L,\uparrow}^{(-)} p_{\uparrow}  + 2 \Gamma_{L,\uparrow}^{(+)}\overline{p}_0^{\uparrow} -  2 \Gamma_{L,\uparrow}^{(-)} \overline{p}^{\uparrow}_{\uparrow} + \Gamma_{L,\downarrow}^{(+)} p_0 +\Gamma_{L,\downarrow}^{(-)} p_{\downarrow}  + 2 \Gamma_{L,\downarrow}^{(+)}\overline{p}_0^{\downarrow} -  2 \Gamma_{L,\downarrow}^{(-)} \overline{p}^{\downarrow}_{\downarrow} \\
& - &  2 ( \Gamma_{L,\uparrow}^{(+)} \overline{p}^{\downarrow}_0 - \Gamma_{L,\uparrow}^{(-)} \overline{p}^{\downarrow}_{\uparrow}) -  2 ( \Gamma_{L,\downarrow}^{(+)} \overline{p}^{\uparrow}_0 - \Gamma_{L,\downarrow}^{(-)} \overline{p}^{\uparrow}_{\downarrow}) - 2 \left[ \frac {} {} (\Gamma_{L,\uparrow}^{(+)} - \Gamma_{L,\downarrow}^{(+)}) p_0 - \Gamma_{L,\uparrow}^{(-)} p_{\uparrow}+ \Gamma_{L,\downarrow}^{(-)} p_{\downarrow} \right]\left( \overline{p}^{\uparrow}_0+  \overline{p}^{\uparrow}_{\uparrow}+ \overline{p}^{\uparrow}_{\downarrow} - \overline{p}^{\downarrow}_0 -  \overline{p}^{\downarrow}_{\uparrow} -  \overline{p}^{\downarrow}_{\downarrow} \right) \nonumber
\end{eqnarray}
With our conventions $(p_0,p_{\uparrow},p_{\downarrow}, \overline{p}^{\uparrow}_0,\overline{p}^{\uparrow}_{\uparrow},\overline{p}^{\uparrow}_{\downarrow},\overline{p}^{\downarrow}_0,\overline{p}^{\downarrow}_{\uparrow},\overline{p}^{\downarrow}_{\downarrow}) \equiv (x_1,x_2,x_3,x_4,x_5,x_6,x_7,x_8,x_9)$, corresponding to an alphabetical ordering $(0,\uparrow,\downarrow)$ where the superscript comes first.

For the practical implementation of the optimal control routines, it is useful to write the above quantities in terms of the coordinates of the  $\mathbf{x}$ vector representing the instantaneous state in the enlarged vector space:
\begin{eqnarray}
i_{\Delta}(t) & = & (\Gamma_{L,\uparrow}^{(+)} - \Gamma_{L,\downarrow}^{(+)}) x_1 - \Gamma_{L,\uparrow}^{(-)} x_2+ \Gamma_{L,\downarrow}^{(-)} x_3 \nonumber \\
i_{\Sigma}(t) & = & (\Gamma_{L,\uparrow}^{(+)} + \Gamma_{L,\downarrow}^{(+)}) x_1 - \Gamma_{L,\uparrow}^{(-)} x_2 - \Gamma_{L,\downarrow}^{(-)} x_3 \nonumber \\
 s_{\uparrow \downarrow}(t)  &  = & \Gamma_{L,\uparrow}^{(+)} x_1 +\Gamma_{L,\uparrow}^{(-)} x_2  + 2 \Gamma_{L,\uparrow}^{(+)} x_4 -  2 \Gamma_{L,\uparrow}^{(-)} x_5 + \Gamma_{L,\downarrow}^{(+)} x_1 +\Gamma_{L,\downarrow}^{(-)} x_3  + 2 \Gamma_{L,\downarrow}^{(+)} x_7 -  2 \Gamma_{L,\downarrow}^{(-)} x_9 \\
& - &  2 ( \Gamma_{L,\uparrow}^{(+)} x_7 - \Gamma_{L,\uparrow}^{(-)} x_8 ) -  2 ( \Gamma_{L,\downarrow}^{(+)} x_4 - \Gamma_{L,\downarrow}^{(-)} x_6) - 2 \left[ \frac {} {} (\Gamma_{L,\uparrow}^{(+)} - \Gamma_{L,\downarrow}^{(+)}) x_1 - \Gamma_{L,\uparrow}^{(-)} x_2+ \Gamma_{L,\downarrow}^{(-)} x_3 \right]\left( x_4+x_5+x_6-x_7-x_8-x_9  \right) \nonumber
\end{eqnarray}


We can derive a similar expression for the fluctuation current associated to the charge transfer by writing
\begin{eqnarray}
 s_{N}(t) & = & \partial_t \left[ \langle (N_{\uparrow} +N_{\downarrow})^2 \rangle - \langle N_{\uparrow} +N_{\downarrow} \rangle^2 \right] \nonumber \\
  &  = &  \langle \mathbf{q}_0 | \left( \partial_{i \chi_{\uparrow}}+\partial_{i \chi_{\downarrow}} \right)^2 \left[ \mathcal{L}(\boldsymbol{\chi},t) | \mathbf{p}(\boldsymbol{\chi},t)  \rangle \right]  |_{\boldsymbol{\chi}=\mathbf{0}} -  2 i_{\Sigma}(t) \langle \mathbf{q}_0 | \left(| \overline{\mathbf{p}(t)}^{\uparrow} \rangle +|\overline{\mathbf{p}(t)}^{\downarrow} \rangle \right) \nonumber \\
    &  = &   \langle \mathbf{q}_0 | \left(\hat{J}_{2,\uparrow} (t)+ \hat{J}_{2,\downarrow} (t) \right) |  \mathbf{p}(t) \rangle + 2 \langle \mathbf{q}_0 | \hat{J}_{1,\uparrow}  (t) | \overline{\mathbf{p}(t)}^{\uparrow} \rangle + 2 \langle \mathbf{q}_0 | \hat{J}_{1,\downarrow} (t) | \overline{\mathbf{p}(t)}^{\downarrow} \rangle  \nonumber \\
& + &    2 \langle \mathbf{q}_0 | \hat{J}_{1,\uparrow}(t) | \overline{\mathbf{p}(t)}^{\downarrow} \rangle +  2 \langle \mathbf{q}_0 | \hat{J}_{1,\downarrow} (t) | \overline{\mathbf{p}(t)}^{\uparrow} \rangle -  2 i_{\Sigma}(t) \langle \mathbf{q}_0 |\left(|\overline{\mathbf{p}(t)}^{\uparrow} \rangle+|\overline{\mathbf{p}(t)}^{\downarrow}  \rangle \right) 
\end{eqnarray}
so that in terms of the coordinates, one finds:
\begin{eqnarray}
s_{N}(t)  &  = & \Gamma_{L,\uparrow}^{(+)} x_1 +\Gamma_{L,\uparrow}^{(-)} x_2  + 2 \Gamma_{L,\uparrow}^{(+)} x_4 -  2 \Gamma_{L,\uparrow}^{(-)} x_5 + \Gamma_{L,\downarrow}^{(+)} x_1 +\Gamma_{L,\downarrow}^{(-)} x_3  + 2 \Gamma_{L,\downarrow}^{(+)} x_7 -  2 \Gamma_{L,\downarrow}^{(-)} x_9 \nonumber \\
& + &  2 ( \Gamma_{L,\uparrow}^{(+)} x_7 - \Gamma_{L,\uparrow}^{(-)} x_8 ) +  2 ( \Gamma_{L,\downarrow}^{(+)} x_4 - \Gamma_{L,\downarrow}^{(-)} x_6) - 2 \left[ \frac {} {} (\Gamma_{L,\uparrow}^{(+)} + \Gamma_{L,\downarrow}^{(+)}) x_1 - \Gamma_{L,\uparrow}^{(-)} x_2- \Gamma_{L,\downarrow}^{(-)} x_3 \right]\left( x_4+x_5+x_6+x_7+x_8+x_9  \right) \nonumber \\
\end{eqnarray}

\section*{S2: Reduction of the correlation between spin and charge FCS}
\label{SM:FCS-spin-transport }

Here, we apply the method described in the main text to uncorrelate the charge and spin fluctuations. To this end, we consider the cost function $ f_0(t)= w_1 s_{N}(t)+ w_2  s_{\uparrow \downarrow}(t)$ with $w_1>0$ and $w_2<0$  in order to maximize the difference $|\Delta \tilde{S}_T^2-\Delta N_T^2|$~(here $\Delta \tilde{S}_{T,\rm ini} > \Delta N_{T,\rm ini}$). Figure~\ref{fig:PlotSeparateFluctuations} shows that the spin and charge fluctuations are driven in opposite directions by the optimization routine. Our optimized cycle generates respectively spin and charge fluctuations $\Delta \tilde{S}^2_{T,\rm opt}=0.153$ and $\Delta N^2_{T,\rm opt}=0.087$ against $\Delta \tilde{S}^2_{T,\rm ini}=0.124$ and $\Delta N^2_{T,\rm ini}=0.099$ for the initial cycle. The ratio $\Delta \tilde{S}^2_{T}/\Delta N^2_{T}$ is thus increased by $25\%$ through the optimization process. Figure \ref{fig:PlotSeparateFluctuations} also presents the optimal driving parameters $\{ V_L(t/T),V_R(t/T) \}$~(the control procedure yields very small values for the magnetic couplings).

\begin{figure}[htbp!]
\begin{center}
\includegraphics[width=8cm]{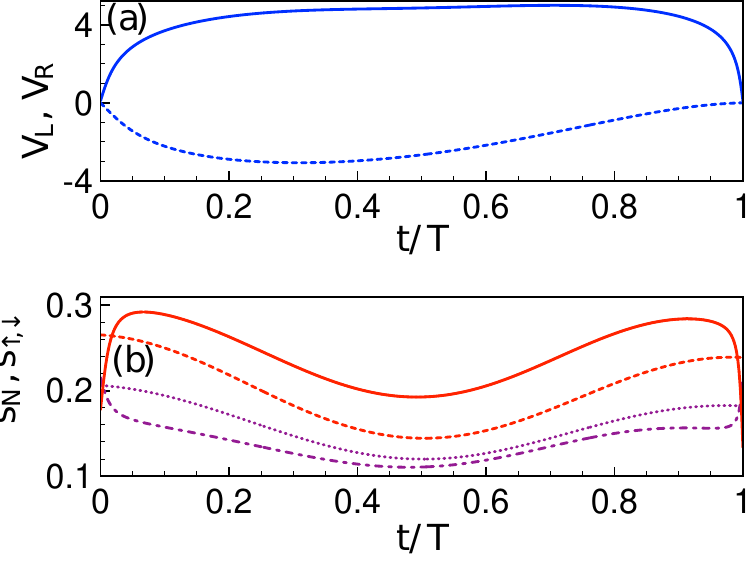}
\end{center}  \caption{Separate driving of the spin and charge fluctuations $\Delta S^2$ and $\Delta N^2$ after $100$ iterations: (a) Evolution of the control parameters $V_L(t/T)$~(solid blue line), $V_R(t/T)$~(dashed blue line) as a function of the rescaled time $t/T.$ (b) Dimensionless spin ($s_{\uparrow,\downarrow}(t/T)$)~(solid red line) and charge ($s_{N}(t/T)$)~(dash-dotted purple line) noise currents as a function of the rescaled time for the optimized cycle. Dashed red and dotted purple lines represent respectively the spin and charge fluctuations for the initial cycle. Parameters: $\Omega=50 \Gamma_0$, increment $\epsilon(t)=4 V_0 \sin(\pi t/T)$, $w_1=2$, $w_2=-2$, $\epsilon=4$. Prescribed cycle $\{ \Gamma_L(t), \Gamma_R(t) \}$, and initial profiles for $\{ V_{L,R}(t/T),\epsilon_{L,R Z}(t/T) \}$ identical to Fig.3 of the main text.}
 \label{fig:PlotSeparateFluctuations} 
\end{figure}

\end{widetext}


\end{document}